\documentstyle[12pt]{article}
\begin{document}
\setlength{\topmargin}{-1cm}
\setlength{\oddsidemargin}{0cm}
\setlength{\evensidemargin}{0cm}
\title{
\begin{flushright}
{\large \bf CERN-TH/96-11\\ MPI-PhT/96-4}
\end{flushright}
\vspace{1cm}
 {\Large\bf Naturally Light Higgs Doublet in the Spinor Representation of 
SUSY $SO(10)$}}

\author{{\bf Gia Dvali}\thanks{E-mail:
dvali@surya11.cern.ch}\\ CERN, CH-1211 Geneva 23, Switzerland\\
\and
{\bf Stefan Pokorski}\thanks{On  leave of absence from Institute
for Theoretical Physics, Warsaw 
University.}\\ Max-Planck-Institute f\"{u}r Physik,
F\"{o}hringer Ring 6, 80805 \\Munich, Germany\\}
 
\date{ }
\maketitle
 
\begin{abstract}
 Within the supersymmetric $SO(10)$ GUT we explore the possibility of the
light Higgs doublet being a member of the $16$-dimensional spinorial
representation. This fact is ultimately related with the assumption that
the light matter (at least partially) resides in some of the tensor
representations as well. Several interesting features emerge. First,
provided that the same $16$-plet is responsible for the breaking
$SO(10) \rightarrow SU(5)$, the heaviness of the top can automatically
follow from the field content at $M_{GUT}$, without need of
any flavour symmetries. Secondly, the doublet--triplet splitting problem
receives a new natural solution. In addition, a dimension=5 (Higgsino
mediated) proton decay can be naturally suppressed. 
We construct explicit $SO(10)$ models with the
above properties, with most general superpotentials under the symmetries.
\end{abstract}
 
\newpage
 
\subsection*{1. Introduction}
 
 The $SO(10)$ group has  many beautiful properties which make it a 
promising candidate for a realistic grand unified theory (GUT). As is
known, the  minimal version of this theory is very constrained:
each family of quarks and leptons belongs
to a single spinorial irreducible representation $16^{\alpha}$-plet (where
$\alpha = 1,2,3$ stands for the family index) and the pair of light
electroweak Higgs doublets resides in a single $10$-plet. 
This which
should lead this theory to successful predictions, unfortunately,
is precisely what makes it incompatible with experiment.
The reason is that the standard Yukawa coupling
\begin{equation}
    Y_{\alpha, \beta}16^{\alpha} 16^{\beta}10,
\end{equation}
`knows' only about the breaking of the electroweak symmetry (through the
vacuum expectation value (VEV) of the $10$-plet)
and also about the `up--down'
symmetry breaking through the well-known parameter tan$\beta$ (ratio
of the `up' and `down' Higgs doublet VEVs). For the heaviest family
this offers the exciting possibility of
top--bottom Yukawa unification for the
large tan$\beta$ regime\cite{tb}. However, some  prediction for the first
two generations
(e.g. $m_\mu = m_s$ and $m_c/m_t = m_s/m_b$) are a disaster.
To avoid the problem, the light-fermion Yukawa couplings should receive
the message about
the GUT symmetry breaking at the tree level.
This naturally calls for the assumption of
the light matter residing not purely in three $16$-plets, but also (at least
partially) in some vector-like representations of $SO(10)$, which can
directly couple to the GUT breaking VEVs in the superpotential \cite{ex}.
There is no reason to assume that this vector-like matter should
include only new $16,\overline{16}$ pairs
and the tensor representations
such as $45$-plet can naturally be there as well. The latter possibility
automatically implies the light Higgs doublet(s) residing in the
spinorial Higgs representation(s). The simplest candidate for such a Higgs
is $16, \overline {16}$ pair, which at the
same time is the simplest candidate for
the breaking of $SO(10)$ down to $SU(5)$ and is automatically present in the
minimal schemes.
Of course, the $16$-plet Higgs in which the light doublets reside, may not
be the same as induces the breaking $SO(10)\rightarrow SU(5)$.
The possibility for the light Higgs doublet to reside in
$16$ was usually ignored in the literature
(for some exceptions see, e.g. \cite{16}), and is the subject of
the present paper. We identify at least two interesting aspects of this
scenario.
Especially interesting we find the
situation where there is a single $16, \overline {16}$ pair in the theory
which at the same time does the GUT symmetry breaking and delivers a
pair of the light doublets. This offers an interesting possibility to
explain the heaviness of the top quark in
terms of the $SO(10)$ representation
content of the matter multiplets, which are light above the GUT scale.
 
We construct a simple complete model of this kind in which the light Higgs
doublets are also partially contained in some 10-plets. The doublet--triplet
splitting is generated inside the 10-plet via the usual `missing VEV
mechanism' \cite{dw} and transmitted to the 16's through the proper mixing. 
The top--bottom
hierarchy automatically results from the hierarchy of GUT breaking scales,
even for the low values of tan$\beta$.
 
However, the scenario with light doublets in $16$ and $\overline {16}$ also
offers a new possibility to solve naturally the doublet--triplet splitting
problem in terms of the symmetries and the group structure and
without any light $10$'s in the spectrum. 
Here also we construct the simplest model.
In this model the pair of $16,\overline {16}$ that contains the
light Higgs doublets is not the same as the one that
breaks $SO(10)$ to $SU(5)$;
therefore, in this model we do not address the question of the heaviness
of the top quark. Although in our simplest models the two new aspects
of the light Higgs doublets in $16$, $\overline {16}$ are not simultaneously
realized, this can in principle be achieved at the expense of a more
complicated heavy spectrum.
 
\subsection*{2. Naturally heavy top}
 
Let us address the issue of the fermion masses in the context
of the light Higgses ($H_u,H_d$) living in the $16$-dimensional
spinor representations $\chi$,$\bar{\chi}$,
which at the same time breaks $SO(10)$
to $SU(5)$.
 As pointed out above, this implies the light matter
residing (at least partially) in some tensor representations.
 
Let us assume for a moment that
this set includes a single $45$-plet with the mass $\sim M_G$, whereas
other states are much heavier $\sim M_{Planck}$ or so
(in particular this can be the case if $45$ transforms under some extra
symmetry broken at $M_G$). Then the relevant Yukawa
couplings are:
\begin{equation}
g_{\alpha}\bar {\chi}16^{\alpha}45 + M_{45}45^2
\end{equation}
where $g_{\alpha}$ are the Yukawa coupling constants and without
loss of generality one can always redefine $16^3=16^{\alpha}g_{\alpha}
(g_{\beta}^2)^{{1 \over 2}}$. Thus $45$-plet couples effectively to a single
$16^3$. Now inserting the $SU(5)$--singlet VEV of $\bar {\chi}$, we find
the following mass matrix of the states that transform as $10$ under $SU(5)$
\begin{equation}
   \overline{10}_{45}[10_{16^3}g\langle \chi \rangle + 10_{45}M_{45}],
\end{equation}
where $g = (g_{\beta}^2)^{1/2}$ and
$\langle \chi \rangle \sim M_{45}$ is an $SU(5)$-singlet VEV. Thus, one
superposition
$10_{light}= [10_{16^3}M_{45} - 10_{45}\langle \chi \rangle]
(M_{45}^2 + g^2\langle \chi \rangle ^2)^{{1 \over 2}}$ stays 
massless at this stage and gets
the tree-level mass after electroweak symmetry breaking from the coupling
\begin{equation}
  5_{\bar{\chi}}10_{light}10_{light}.
\end{equation}
This induces the mass of a single up-type quark only, naturally to be
identified with the top. The rest of the matter fermions are left massless
at this stage. Thus, top is the naturally the 
heaviest state in this picture and
this is achieved without any
flavour symmetry, or other input difference among
families, but just due to a field content at $M_G$\footnote{1 A similar
approach describing the quark mass hierarchy in terms of the heavy field
content at $M_G$ (without appealing to flavour symmetry) was developed in
the context of $SU(6)$ GUT\cite{bdsbh}}

Now what about the masses of the lighter fermions?
Although the explanation of the complete mass pattern will 
not be attempted here, we will briefly mention
some possibilities. One way to
generate the masses of the lighter families is to assume that they come from
the effective operators induced after integrating out the vector-like
matter at scales larger than $M_G$.
 
We thus have to assume couplings
\begin{equation}
\chi16^{\alpha}N + \bar{\chi}16^{\alpha}R + M_RR^2 +M_NN^{2}
\end{equation}
where $N$ and $R$ are some heavy fields in the appropriate tensor
representations contained in the products $16\times16$ and
$16\times \overline{16}$. $M_N$ and $M_R$ have to be understood as certain
superpositions of constant mass terms with the GUT-breaking Higgs VEVs.
After integrating them out,
we will get
effective operators of the form:
\begin{equation}
{1 \over M_N}\chi\chi16^{\alpha}16^{\beta} +
{1 \over M_R}\bar{\chi}\bar{\chi}16^{\alpha}16^{\beta}
\end{equation}
which induce the masses of the light fermions.
An alternative way is to assume
that the small admixture ($\sim \epsilon$) of the light doublet resides
also in $10'$-plet Higgs.
Then the lighter masses can be induced from the coupling
\begin{equation}
10'16^{\alpha}16^{\beta}.
\end{equation}
In the next section we show that the smallness of $\epsilon$ and thus the
top--bottom mass hierarchy can be directly related to the hierarchy
of the $SO(10)$-breaking scales and namely $\chi/A \sim$ tan$\beta m_b/m_t$,
where $\chi$ and $A$ are two VEVs invariant under $SU(5)$ and
$SU(4)_C\otimes SU(2)_L\otimes U(1)_R$ respectively.
 
\subsection*{3. A complete model with naturally heavy top}

We begin with the discussion of the doublet--triplet splitting mechanism.
As already mentioned, in our model the light Higgs doublets
are partially located in the same
$16$ that breaks $SO(10)$ to $SU(5)$
(to get a natural solution for the heaviness of the top quark)
and partially in some $10$-plet
(depending on the symmetries this $10$-plet may or may
not be coupled to light
fermions).
A realistic model requires the existence of a
$45$-plet Higgs (hereafter denoted by $A$) with the VEV of the form
\begin{equation}
\langle A \rangle = diag[0,0,0,A,A]\otimes \epsilon
\end{equation}
where $A \sim M_{GUT}$, and each element is assumed to be proportional
to the $2 \times 2$ antisymmetric matrix $\epsilon$. This  VEV breaks
the $G_{L,R} = SU(2)_L\otimes SU(2)_R \otimes SU(4)$ subgroup of $SO(10)$
down to $SU(2)_L\otimes U(1)_R \otimes SU(4)$ and is thus oriented along 
the $T_R^3$ generator of $SU(2)_R$.  In combination with the other VEVs,
say the $16$-plet with non-zero SU(5)-singlet VEV, it leads
to the desired breaking $SO(10) \rightarrow G_W =SU(3)_c\otimes SU(2)_L
\otimes U(1)_Y$.

We also need another 45-plet Higgs (to be denoted by $B$) with the VEV
along the $B$--$L$ direction
\begin{equation}
\langle B \rangle = diag[B,B,B,0,0]\otimes \epsilon
\end{equation}
in order to ensure the doublet--triplet splitting inside the $10$-plet
via the usual  `missing VEV' mechanism \cite{dw}. From the first glance the
introduction of the second $45$-plet looks somewhat unmotivated, since
the group theoretically single $45$-plet $A$, in combination with
$\chi, \bar {\chi}$, is quite enough to break the $SO(10)$ group down to
$G_W = SU(3)_C\otimes SU(2)_L\otimes U(1)$. However,
on practice it turns out that the most general renormalizable superpotential
of single $45$ and $16,\overline 16$ pair does not allow for such a
minimum (the only sensible minimum is the $SU(5)$-symmetric one).
Thus, the introduction of extra $45$-plet(s) (and even $54$-plet) is
necessary in order to complete the breaking (see section 6).
In this section we
construct the superpotential which is most general under symmetries and
automatically delivers a desired VEV structure. Now let us show
how the above VEV structure creates a light doublet in the same $16$-plet
as  does the GUT symmetry breaking. Consider the superpotential
\begin{equation}
W_{DT} = {g \over 2}\chi\chi 10 + {g' \over 2}\bar{\chi}\bar{\chi} 10 
+ fA1010' + qB10'10''+ {p \over 2}\nu 10''^2
\end{equation}
where $10,10'$ and $10''$ are three $10$-plets and $\nu$ is a gauge singlet
with the VEV $\sim M_G$. As shown below the
above structure can be a natural
consequence of an exact symmetry of the theory and the three 10-plets
is a minimal set needed for our purposes (see eq(11)).
In the full theory, the above superpotential has
to be supplemented with the  piece that includes the interaction of
$\chi,\bar{\chi}$ with other GUT Higgses and induces its VEV in the
$SU(5)$ singlet direction. This will be done below, but now let us ignore it
for the moment and only remember that the components of $\chi, \bar{\chi}$
that transform as $10, \overline{10}$ under $SU(5)$ become heavy
(1) because they partially reside in the Goldstone multiplets that are
eaten up by the gauge superfields of broken generators $SO(10)/SU(5)$;
and (2) because of
the mixing with the similar components from Higgs $45$-plets.
Thus, we have to discuss only the masses of the states that transform
as $5,\bar {5}$ of $SU(5)$. It is not difficult to see that the
mass matrices of the doublet and triplet components from $10,10',10'',
\chi, \bar{\chi}$ have the following form:
\begin{eqnarray}
M_{doublet} = \left(\begin{array}{cccc}
0 & g\chi & 0 & 0\\
g'\chi & 0 & fA & 0\\ 
0 & -fA & 0 & 0\\
0 & 0 & 0 & p\nu \end{array} \right);~ 
M_{triplet}=\left(\begin{array}{cccc}
0 & g\chi & 0 & 0 \\
g'\chi & 0 & 0 & 0\\
0 & 0 & 0 & qB\\
0 & 0 & -qB & p\nu \end{array} \right).
\end{eqnarray}

Thus, the triplets are all heavy, whereas the doublet mass
matrixes have one zero eigenvalue each:
\begin{equation}
H_{light} = [H_{\chi}A - H_{10'}g'\langle \chi \rangle]
(g^2\langle \chi \rangle^2 + A^2)^{1/2}
\end{equation}
Thus, the admixture of the light doublet(s) in $\chi, \bar{\chi}$
and $10'$ is controlled by the hierarchy of the symmetry-breaking
scales $SO(10)\rightarrow SU(4)\otimes SU(2)_L\otimes U(1)$
versus $SO(10)\rightarrow SU(5)$. Say, if the latter scale is somewhat
smaller, the light doublet will predominantly live in $\chi,\bar{\chi}$.
As was argued above, this offers an interesting possibility for
relating the top--bottom mass hierarchy to the hierarchy of the
symmetry breaking.
 
We are now in a position to write the complete
superpotential of the theory.
In fact what we need now is to take care of the Higgs
sector that breaks GUT symmetry and satisfies the following requirements:
 
 ($a$) $W_{Higgs}$ should be most general under symmetries;
 
 ($b$) there should be no `fine-tuning';
 
 ($c$) it should allow for the $G_w$-symmetric SUSY minimum in which
the only light states are the ones of the minimal supersymmetric
standard model (+ possibly some $G_W$-singlets) and
a pair of light doublets residing at least partially in Higgs $16$-plets.

The Higgs superpotential includes the chiral superfields in the
following $SO(10)$ representations:
$S,X,Y$-singlets; $\Sigma$-$54$-plet, $A,B,C$-$45$-plets;
$\chi, \bar {\chi}$ - $16,\overline {16}$-plets and
three $10$-plets $10, 10',10''$.
We also introduce two singlets $\nu, \nu'$ and the one
$45$-plet in the matter sector.
 
The Higgs superpotential has the form:
\begin{equation}
W_{Higgs} = W_{GUT} + W_{DT}.
\end{equation}
The two parts are given as
\begin{eqnarray}
W_{GUT} &=& {\sigma \over 4}STr\Sigma^2 + {h \over 6} Tr\Sigma^3 +
{1 \over 4}Tr(a\Sigma + M_a + a'S)A^2 +
{1 \over 4}Tr(b\Sigma + M_b + b'S)B^2 \nonumber\\
&+& {1 \over 2}Tr(a''XA + b''YB)C + {g_c \over 2}\bar {\chi}C\chi +
(r_3\nu + r_4\nu')TrC^2 + M^2S \nonumber\\
&+& {M' \over 2}S^2 + {\kappa \over 3}S^3
\end{eqnarray}
and
\begin{equation}
W_{DT} = g\chi 10 \chi + \bar {g}\bar {\chi} 10\bar {\chi} +
 fA10'10 + qB10'10'' + r\nu10^2 + (r_1\nu + r_1\nu')10^{''2}
\end{equation}
This form is strictly natural, since it is the most general compatible
with the $Z_2^A\otimes Z_2^B \otimes U(1)^C$ global symmetry under
which the chiral superfields  transform as follows:
 under $Z_2^A$
\begin{equation}
 (A,X,10',10'') \rightarrow -(A,X,10',10'')
 \end{equation}
 under $Z_2^B$
\begin{equation}
(B,Y,10'') \rightarrow -(B,Y'10'')
\end{equation}
and under $U(1)^C$
\begin{eqnarray}
 (C,10,10'') &\rightarrow& e^{i2\alpha} (C,10,10'')\nonumber\\
 (\chi,\bar {\chi}) &\rightarrow& e^{-i\alpha}(\chi,\bar {\chi})\nonumber\\
 (X,Y,10') &\rightarrow& e^{-i2\alpha} (X,Y'10')\nonumber\\
 16^{\beta} &\rightarrow& e^{i\alpha} 16^{\beta}\nonumber\\
(\nu, \nu') &\rightarrow& e^{-i4\alpha} (\nu, \nu').
\end{eqnarray}
 
Besides we also assume `matter parity' under which $16^{\alpha}$ and
$45$ change sign and all other superfields are invariant.
 
The above superpotential admits
the following supersymmetric ($F$-flat and $D$-flat) minimum with
an unbroken $G_W$ symmetry:
\begin{eqnarray}
\Sigma &=& diag(2,2,2,2,2,2,-3,-3,-3,-3)\Sigma~~~~where~
\Sigma = {b'M_a - a'M_b \over 3ab' + 2ba'}\nonumber\\
A &=& diag[0,0,0,A,A]\otimes \epsilon\nonumber\\
B &=& diag[B,B,B,0,0]\otimes \epsilon\nonumber\\
\chi &=& \bar {\chi} = \chi|+,+,+,+,+\rangle~~
where~\chi^2 = -{a'' \over g_c} XA = -{b'' \over g_c} YB\nonumber\\
S &=& -{2bM_a + 3aM_b \over 3ab' + 2ba'}\nonumber\\
 10 &=& 10' = 10'' = C = \nu = 0
\end{eqnarray}
 
According to the standard notations (e.g. see \cite{wz}) the
$SU(5)$ singlet component of $16$ is denoted
by $|+,+,+,+,+\rangle$, where each $`+'$
refers to an eigenvalue
of the respective Cartan subalgebra generator.
The two quantities $A$ and $B$ are determined
from the two equations:
\begin{eqnarray}
&& 10(S\sigma \Sigma - h\Sigma^2) - aA^2 + bB^2 =0\nonumber\\
&& 15\sigma\Sigma^2 + a'A^2 + {3 \over 2}b'B^2 + M^2 +M'S +
\kappa S^2 =0
\end{eqnarray}
Note that the absolute VEVs of the singlets $X$ and $Y$ are undetermined
in the SUSY limit, and only their ratio, 
${X \over Y} = {b''B \over a''A}$, is determined.
The VEV of $\nu'$ is also undetermined in the SUSY limit
and we assume that it will be fixed $\sim M_G$ after
supersymmetry breaking.
Then, according to the analysis of section 2, the light doublet pair
is partially residing in $\chi, \bar{\chi}$ and $10'$ representations,
and its admixture is controlled by the hierarchy of the scales
($\langle \chi \rangle/A$ ratio). 
Since the $\langle \chi \rangle$ VEV parametrizes a SUSY flat
direction, in the SUSY limit the theory admits two realistic vacuum states:
in the limit $\langle \chi \rangle >> A$ it recovers an `old' 
top--bottom unification
prediction of the minimal $SO(10)$, whereas in the other case,
$\langle \chi \rangle << A$, it leads to the heavy top for the low
tan$\beta$ regime.

\subsection*{4. Higgsino-induced proton decay}

Considered scheme allows for the natural suppression of the 
coloured Higgsino-mediated (dimension=5) proton decay\cite{d5}, which is 
usually a problem in the standard approaches
(for some alternative solutions
within the $SO(10)$ GUT see, e.g.\cite{dbb}). As it is known the dangerous
dimension=5 operators can occur only if there is a supersymmetric
(chirality flip) mass insertion $M_T\bar TT$ which mixes the colored
triplet partners $T,\bar T$ of $H_u$ and $H_d$ Higgs doublets respectively.
To study suppression we have to identify $T,\bar T$ states that are
coupled to light matter fermions. This depends on the way the
light family masses are generated and, therefore,
we separately discuss two above-mentioned possibilities.

 One possibility is that $10'$ is decoupled from the quarks and leptons
and the masses are generated from the couplings (4) and (6) (which are
low energy remnant of (2) and (5) respectively). In this case 
the only colored triplets coupled to light matter are 
$\bar T_{\chi}$ and $T_{\bar {\chi}}$ states from $\chi$ and $\bar{\chi}$
respectively. Due to (11) these triplets have no supersymmetric
crossing mass term and thus, there are no Higgsino induced 
dimension=5 operators.

Now assume that at least some of
the light fermion masses are induced from the coupling
(7). Then, there is another pair of coloured triplets
$T_{10'}, \bar T_{10'}$ coupled to the quark and lepton superfields.
However, as it can be seen from (11), their chirality flip mass
term is only induced due to their mixing with $10''$ (and not directly 
with each other) and, therefore, for $\nu << qB$ it is suppressed
by an additional factor $\sim {\nu \over qB}$. Presumably we can not take
$\nu$ arbitrarily small (since it measures a mass of the
doublet component from $10''$ and this in general can affect the
unification of gauge couplings), but even assuming it one or two orders
of magnitude below $M_G$ one can significantly improve the situation
with respect to a standard case.  
 
\subsection*{5. A model for the doublet--triplet splitting without
a light $10$-plet}
 
In the previous section we have constructed a simple model with
emphasis on the naturally heavy top quark. This is achieved by requiring
the light Higgs doublet to be in the same 16 as breaks $SO(10)$.
To obtain doublet--triplet splitting in this model with only
one pair $16, \overline {16}$, the simplest possibility is to introduce
also a $10$-plet. The doublet--triplet splitting in the 10-plet is
transmitted to the 16-plet and the light Higgs doublet is partially in 16
and partially in 10. The mixing is controlled by $A/\langle \chi
\rangle$ which at
the same time can give heavy top even for low tan$\beta$. In this
section we put aside the problem of the heavy family quark masses
(we assume it can be solved by one of the known (or unknown!)
mechanisms) and construct an explicit model where the ansatz about
light Higgs doublet in $16,\overline {16}$ is used for a non-standard
solution to the doublet--triplet splitting. We propose a mechanism
that makes use solely of 16-plets. One may ask what are the motivations
for avoiding light doublets residing in the $10$-plet Higgs.
First, the scenario without light $10$-plets is logically different
possibility which is certainly worth studying. Secondly, we think
that the introduction of a heavy vector-like matter (which seems in
any case is necessary for the avoiding the wrong mass relations) 
in this scenario looks more motivated than in the standard versions
(with light doublets in $10$-plet): in the former case quark and
lepton masses simply can not be generated without exchange of heavy
vector-like states (see eq(5)), whereas in the later case they can be
induced easily.

For the Higgs spectrum we take again a 45-plet $A$ with the $T_{3R}$
VEV, a pair $\chi, \bar{\chi}$ of $16,\overline {16}$ which
breaks $SO(10)$, a pair $\psi, \bar {\psi}$ of $16,\overline {16}$ with
light Higgs doublet and another heavy Higgs $45$-plet $\Phi$.

Then, the piece of the superpotential responsible for the doublet--triplet
splitting has the form:
\begin{equation}
W_{DT} = g_a\bar {\psi}A \psi + g_{\Phi} \chi \Phi \bar{\psi} +
 \bar {g_{\Phi}}\bar {\chi} \Phi \psi + MTr\Phi^2.
\end{equation}

The doublet--triplet splitting
mechanism can be easily understood in terms of the $G_{LR}$ invariant
decomposition of $45$ and $16$ representations
\begin{eqnarray}
16 &=& (2,1,4) + (1,2,\bar 4)\nonumber\\
45 &=& (1,3,1) + \dots
\end{eqnarray}
Now, since the VEV of $A$ transforms as $(1,3,1)$ it can give masses
only to the right-hand doublet states in $\psi, \bar {\psi}$. Thus,
all the 
left-handed states ($(2,1,4)$ fragment) stay massless at these stage.
This are the states with quantum numbers of Higgs doublets $H,\bar H$
and left-handed quark doublets $Q,\bar Q$. However,
$Q, \bar Q$ states transform as $10, \overline {10}$ under $SU(5)$
and therefore they are mixed with the similar states from $\Phi$ through
the $\chi, \bar {\chi}$ VEV and become heavy.
In contrast, the $H, \bar H$ have no partners in $\Phi$ (which includes no
component transforming as $5,\bar 5$ under $SU(5)$) and thus stay massless.

We now proceed to construct an explicit model with the proper GUT
potential.
The Higgs superpotential includes the chiral superfields in the following
$SO(10)$ representations: $ S,X,Y \equiv$ singlets;
$\Sigma \equiv 54$-plet; $A,B,C,\Phi \equiv$ 45-plets;
$\chi, \bar {\chi}, \psi, \bar {\psi} \equiv 16,\overline{16}$-plets
and $F \equiv 10$-plet.
The superpotential has the form:
\begin{equation}
W_{Higgs} = W_{GUT} + W_{DT}
\end{equation}
where
\begin{eqnarray}
W_{GUT} &=& {\sigma \over 4}STr\Sigma^2 + {h \over 6} Tr\Sigma^3 +
{1 \over 4}Tr(a\Sigma + M_a + a'S)A^2 +
{1 \over 4}Tr(b\Sigma + M_b + b'S)B^2 \nonumber\\
&+& {1 \over 2}Tr(a''XA + b''YB)C + {g_c \over 2}\bar {\chi}C\chi +
M^2S + {M' \over 2}S^2 + {\kappa \over 3}S^3
\end{eqnarray}
and
\begin{equation}
W_{DT} = g_f\chi F \chi + \bar {g_f}\bar {\chi} F\bar {\chi} +
g_{\Phi}\bar {\psi} \Phi \chi
+ \bar {g_{\Phi}}\bar {\chi} \Phi \psi + g_a\bar {\psi}A \psi
+ \rho X Tr\Phi^2
\end{equation}
Again, this form is strictly natural, since
it is the most general that is compatible
with the $Z_4^A\otimes Z_2^B \otimes U(1)^C$ global symmetry under
which the chiral superfields  transform as follows:
under $Z_4^A$
\begin{eqnarray}
 && (A,X) \rightarrow -(A,X)\nonumber\\
 && (\psi,\bar {\psi}) \rightarrow  i (\psi,\bar {\psi})\nonumber\\
 && \Phi \rightarrow -i\Phi
\end{eqnarray}
 under $Z_2^B$
\begin{equation}
(B,Y) \rightarrow -(B,Y)
\end{equation}
and under $U(1)^C$
\begin{eqnarray}
 (C,F) &\rightarrow& e^{i2\alpha} (C,F)\nonumber\\
 (\chi,\bar {\chi}) &\rightarrow& e^{-i\alpha}(\chi,\bar {\chi})\nonumber\\
 (X,Y) &\rightarrow& e^{-i2\alpha} (X,Y)\nonumber\\
 \Phi &\rightarrow& e^{i\alpha} \Phi
\end{eqnarray}

We assume that all mass scales in $W_{GUT}$
are $\sim M_{GUT}$ and all coupling constants are of the order of 1.

The standard procedure shows that the above superpotential admits
the following supersymmetric ($F$-flat and $D$-flat) minimum with
an unbroken $G_W$ symmetry:
\begin{eqnarray}
\Sigma &=& diag(2,2,2,2,2,2,-3,-3,-3,-3)\Sigma~~~~where~
\Sigma = {b'M_a - a'M_b \over 3ab' + 2ba'}\nonumber\\
A &=& diag[0,0,0,A,A]\otimes \epsilon\nonumber\\
B &=& diag[B,B,B,0,0]\otimes \epsilon\nonumber\\
\chi &=& \bar {\chi} = \chi|+,+,+,+,+\rangle~~
where~\chi^2 = -{a'' \over g_c} XA = -{b'' \over g_c} YB\nonumber\\
S &=& -{2bM_a + 3aM_b \over 3ab' + 2ba'}\nonumber\\
\psi &=& \bar {\psi} = F = \Phi = C = 0
\end{eqnarray}

  As was shown above, in the given vacuum $W_{GUT}$
delivers a pair of the light doublets
from the $\psi, \bar {\psi}$ multiplets. Once again, this is because
$\psi, \bar {\psi}$ states get their masses from only the two sources:
through the VEV of $A$ and via mixing with the heavy $\Phi$ through
the $\chi,\bar {\chi}$ VEV. 
Now, the $A$ VEV leaves all $SU(2)_L$-doublet states in
$\psi,\bar {\psi}$ massless. This is because all these states are zero
eigenvalues of the $T_{3R}$ generator.
These are the states with quantum numbers of
$Q, \bar Q$ and $H, \bar H$.
$Q,\overline Q$ components of $\bar{\psi}, \psi$ are mixed with
the similar components of $\Phi$
through the VEVs of $\chi, \bar {\chi}$ and become heavy.
In contrast, $45$-plet $\Phi$ has no colour singlet weak doublet
component and thus $H, \bar H$ states are massless. All other
$G_W$ non-singlet states from $W_{GUT}$ have a GUT scale mass.
                    
\subsection*{6. Why $W_{GUT}$?}

We have discussed in detail the structure of $W_{DT}$ and $W_{Yukawa}$ 
in the previous sections and each term there plays a well-defined role
in the doublet--triplet splitting mechanism or in the fermion 
mass generation respectively.
They have an obvious property of each term including at least two of
the superfields with a vanishing VEVs in the phenomenologically
acceptable SUSY vacuum. Thus, $W_{DT}$ and $W_{Yukawa}$ cannot affect
the configuration of the VEVs in this vacuum and the
symmetry breaking pattern is solely decided by $W_{GUT}$.
Now what about the structure of $W_{GUT}$? From the first glance its
form looks rather ad hoc and non-minimal. However, the more careful
analysis shows that this form is one of the simplest choices that can lead
to the desired breaking of $SO(10)$ to $G_W$ and deliver the
VEV structure necessary for the doublet--triplet splitting.
In this section we will justify the form of the Higgs superpotential
as being one of the simplest possibilities for the natural solution of
the doublet--triplet splitting problem in $SO(10)$. Below we will
show why the simpler structures do not work. We will also confront the
Higgs sectors needed for the natural doublet--triplet
splitting via two different
mechanisms, with the light doublet appearing in $10$-plets 
(Dimopoulos--Wilczek mechanism)
or $16$-plet. As we know, these mechanisms require
the existence of the $45$-plet Higgs $A$ with the VEV along the $B$-$L$
or $T_{3R}$ directions.
(as discussed in section 3, a combination
of the two is also possible)
and the key point is the existence (among other terms) of the
couplings $A_{B-L}10'10$ or
$A_{T_{3R}}\overline {16}_{Higgs}16_{Higgs}$.

Imagine therefore that the GUT sector includes
the only two representations: a $45$-plet $A$ and a $16,\overline{16}$
pair $\chi, \bar{\chi}$. This set would be a very economical and simplest
choice, since it is sufficient $group~theoretically$ for the desired
symmetry breaking  as well as for the desired VEV structure. Unfortunately,
this simplest possibility does not work, since the $G_W$-symmetric state
never corresponds
to a SUSY minimum of the most general superpotential with an 
arbitrary number
of gauge singlets included
\begin{equation}
W= {1 \over 4}M_ATrA^2 + {1 \over 2}g\bar{\chi}A\chi + M_{\chi}\bar{\chi}\chi
 +(singlet~self-couplings)
\end{equation}

where $M_A$ and $M_{\chi}$ have to be understood as the arbitrary
(up to a symmetries) superpositions of the constant mass terms and the
gauge--singlet superfields. The heart of the problem is in
the $F$-flatness condition for the $45$-plet 
($\sigma_{ki}$ are $SO(10)$ generators in the
spinor representation)
\begin{equation}
F_{A_{ik}} = M_AA_{ki} + g\bar{\chi}\sigma_{ki}\chi = 0,
\end{equation}
which tells us that the only vacuum with
unbroken $G_W$ is an $SU(5)$-symmetric
one. The group theoretical reason for this is that the only acceptable
maximal little group of $16$ is $SU(5)$.
 
The introduction of another $45$-plet $B$ does not help much. To see this,
first notice that a necessary requirement is that an intersection of
two VEVs $A$ and $B$ should
break group to $SU(3)_c\otimes SU(2)_L\otimes U(1)_c
\otimes U(1)_R$, meaning that the matrix
\begin{equation}
\langle A \rangle + \theta \langle B \rangle 
= diag[V,V,V,V',V']\otimes\epsilon
\end{equation}
must have the full diagonal occupied ($ V,V'\neq 0$) for some non-zero
$\theta$. If this condition is not satisfied, the situation is effectively
reduced to the case with a single $45$-plet.
Next, let us point out that no matter whether we
want to split doublet--triplet masses
inside the $10$-plet or $16$-plet Higgs, the
$A$ and $B$ must have different transformation properties
under some symmetry;
otherwise the doublet and triplet mass matrices would have the same
zeros and doublet--triplet splitting would be impossible without fine-tuning.
Thus, only one of the $45$-plets can couple to $\chi, \bar {\chi}$ and we
are lead to the generic superpotential
\begin{equation}
W= {1 \over 4}M_ATrA^2 + {1 \over 4}M_BTrB^2 + {1 \over 2}M_{AB}TrAB + 
{1 \over 2}g\bar{\chi}A\chi + 
M_{\chi}\bar{\chi}\chi + (singlet~self-couplings)
\end{equation}
where, as before, each mass parameter has to be understood as a whatever
combination of singlets and constant mass terms allowed by symmetries that
we will not specify here. This form is enough to point out the problem,
since the $F$-flatness requirement among others implies the following
conditions
\begin{eqnarray}
 F_{A_{ik}}&=& M_AA_{ki} + M_{AB}B_{ki} + g\bar{\chi}\sigma_{ki}\chi = 0
\nonumber\\
F_{B_{ik}} &=& M_BB_{ki} + M_{AB}A_{ki} = 0
\end{eqnarray}
Immediately we see that at best there are two possibilities
(with $G_W$ unbroken):
either $M_B$ and $M_{AB}$ both vanish at the minimum, and thus
$A = SU(5)$-singlet and $B = undetermined$;
or
\begin{equation}
B_{ki} = -{M_{AB} \over M_B} A_{ki} = {gM_{AB} \over M_AM_B - M_{AB}^2}
\bar{\chi}\sigma_{ki}\chi = SU(5)-singlet
\end{equation}
Clearly none of the two is acceptable.
 
The next logical step is to introduce $54$-plet $\Sigma$, which
is the lowest dimensional representation that contains a singlet
of $G_W$, but no singlet of $SU(5)$. Unfortunately, $\Sigma$ can not
substitute one of the $45$-plets, since it cannot couple to the
$16$-plets (there is no $54$ contained in either of the  direct products
$16\otimes 16$ or $\overline {16}\otimes 16$). Therefore, the system
with $54$ ,$45$ and $16,\overline{16}$, although it can successfully break
the $SO(10)$ to $G_W$, can never deliver a desired VEV for $45$-plet.
Thus, $\Sigma$ has to be introduced on top of the
$A,B,\bar{\chi},\chi$
states. For constructing the superpotential, we have to remember
that: (1) as pointed above, $A$ and $B$ must transform differently under
some symmetry, and (2) superpotential must contain a crossing coupling
$Tr\Sigma A B$ or $M_{AB}TrAB$. The second condition is needed in order
to eliminate the unwanted massless 
pseudo-Goldstone states, which otherwise
would be presented since, without a direct cross coupling,
$A$ and $B$ can communicate only through the $SO(6)\otimes SO(4)$-invariant
VEV of $\Sigma$ and, consequently, the correct vacuum would have an
accidental global $[SO(6)\otimes SO(4)]_A\otimes [SO(6)\otimes SO(4)]_B$
degeneracy. Thus, the superpotential should have the following generic
form
\begin{eqnarray}
W &=& M_{\Sigma}Tr\Sigma^2 + hTr\Sigma^3 + Tr[(aA^2 +bB^2 + cAB)\Sigma] +
{1 \over 4}M_ATrA^2 + {1 \over 4}M_BTrB^2 \nonumber\\
&+& {1 \over 2}M_{AB}TrAB +
{1 \over 2}g\bar{\chi}A\chi + M_{\chi}\bar{\chi}\chi +
(singlet~self-couplings)
\end{eqnarray}
where at least one of the two constants $a$ and $b$ should be zero.
It is not difficult to find the combination of the symmetries and the
singlet fields, which allow for the desired symmetry breaking and the
$A$-plet having the VEVs in
the $B$-$L$ or $T_{3R}$ direction (or in both).
However, the problem is that whenever an $SU(5)$-singlet from $\chi$
gets a VEV, the $F$-flatness condition implies
\begin{equation}
M_{\chi} + 3gA = 0~~~~or~~~~  M_{\chi} + 2gA = 0
\end{equation}
for the $A \sim B-L$ and $A \sim T_3R$ cases, respectively. First of all,
this immediately means that we cannot extract the light doublet
from the same $\chi, \bar{\chi}$, because in these two cases
they get the mass equal to $2M_{\chi}$ and $M_{\chi}$, respectively.
However, this is not a major
problem, just the fact that the quantity $M_{\chi}$ at the minimum must be
non-zero. This quantity  must transform
identically to $A$ under any symmetry
other than $SO(10)$ and thus, it is very difficult to prevent a multiplet
in which doublet--triplet
splitting is induced by the VEV of $A$ from having a large
$SO(10)$ invariant mass. One way to improve the situation could be
to assume that $M_{\chi}$ is a superposition of several singlets
which, being coupled to $10'10$ or $\overline{16}_{Higgs}16_{Higgs}$,
exactly compensate each other. Such a solution is possible,
but we think that a much simpler
way is to introduce another $45$-plet $C$ with vanishing VEV
in the right vacuum. Thus, the system with $\Sigma,A,B,C,\chi,\bar{\chi}$
corresponds to the simplest possibility and we are lead to the $W_{GUT}$
discussed in the previous sections.
 
\subsection*{7. Summary}
Model building at the GUT scale is useful if taken with
the proper attitude: as an illustration of certain general possibilities
to solve the existing physical (or technical) problems in the framework
of GUTs. On the other hand, it is not sensible to aim already now at
the `true' theory. In this paper we have shown that light Higgs
doublets in $16,\overline {16}$ of $SO(10)$ is a viable alternative, which
may explain the heaviness of the top quark and offer a new mechanism for
doublet--triplet splitting. We have also shown that it is easy to
construct self-consistent (and relatively simple) models with
these properties.

\subsection*{Acknowledgements}
S.P. was partially supported by the Polish committee for Scientific
Research and by the EU grant Flavourdynamics.

\end{document}